\documentclass[prd,aps,showpacs,nofootinbib,twocolumn,superscriptaddress,
amssymb]{revtex4}

\usepackage{graphicx}
\usepackage[english]{babel}
\usepackage{amsmath}
\usepackage{amssymb}
\usepackage{amsfonts}
\usepackage{latexsym}

\newcommand{\be}{\begin{equation}}
\newcommand{\ee}{\end{equation}}
\newcommand{\bea}{\begin{eqnarray}}
\newcommand{\eea}{\end{eqnarray}}
\newcommand{\beaa}{\begin{eqnarray*}}
\newcommand{\eeaa}{\end{eqnarray*}}

\def\be{\begin{equation}}
\def\ee{\end{equation}}
\def\bea{\begin{eqnarray}}
\def\eea{\end{eqnarray}}

\begin{document}

\title{On the Viability of the Intermediate Inflation Scenario with $F(T)$ Gravity}

\author{V.K. Oikonomou}
\email{v.k.oikonomou1979@gmail.com}
\affiliation{Laboratory for
Theoretical Cosmology, Tomsk State University of Control Systems and
Radioelectronics (TUSUR), 634050 Tomsk, Russia}
\affiliation{Tomsk
State Pedagogical University, 634061 Tomsk, Russia}

\begin{abstract}
We study the intermediate inflation scenario in the context of
$F(T)$ gravity, and we examine it's viability by calculating the
power spectrum of the primordial curvature perturbations and the
corresponding spectral index. As we demonstrate, it is possible for
the resulting spectral index to be compatible with the observational
data and we investigate the parameter space in order to see when the
compatibility with data is possible.
\end{abstract}



\maketitle

\section{Introduction}

An alternative theory to Einstein-Hilbert gravity is $F(T)$ gravity,
where the fundamental quantity that is used is the torsion $T$
corresponding to the Weitzenb\"{o}ck connection
~\cite{Hehl:1976kj,Hayashi:1979qx,Cai:2015emx,Flanagan:2007dc,Maluf:2013gaa,Ferraro:2008ey},
instead of the Ricci scalar corresponding to the Levi-Civita
connection. In general, modified gravity models of teleparallelism,
in which case the theory is built by using a appropriately chosen
function of the torsion $T$, can explain various cosmological eras
of our Universe, for example the late-time acceleration issue in the
context of $F(T)$ gravity was studied in Refs.
\cite{Bamba:2013jqa,Bengochea:2008gz,Linder:2010py,Geng:2011aj,Otalora:2013tba,Chattopadhyay:2012eu,Dent:2011zz,Yang:2010hw,Bamba:2010wb,Capozziello:2011hj,Geng:2011ka,Farajollahi:2011af,Cardone:2012xq,Bahamonde:2015zma},
while inflationary and bouncing scenarios in relation with
cosmological perturbation scenarios were studied in Refs.
\cite{Cai:2011tc,Chen:2010va,Izumi:2012qj,Nashed:2014lva,Hanafy:2014bsa,Hanafy:2014ica,Ferraro:2006jd}.
Moreover, special cosmological solutions where found in Refs.
\cite{Rodrigues:2012qua,Li:2013xea} and special metric solutions or
astrophysical objects studies where performed in Refs.
\cite{Capozziello:2012zj,Paliathanasis:2014iva,Gonzalez:2011dr,Bohmer:2011si,Nashed:2013bfa,Ruggiero:2015oka},
and in addition thermodynamical issues were addressed in
\cite{Bamba:2012vg,Bamba:2010wb,Karami:2012fu}. This research stream
renders $F(T)$ gravity an important alternative to Einstein-Hilbert
gravity. To this end, the purpose of this paper is to investigate if
a quite popular inflation scenario, and specifically the
intermediate inflation scenario
\cite{Barrow:1990td,Barrow:1993zq,Rezazadeh:2014fwa,Barrow:2006dh,Barrow:2014fsa,Herrera:2014mca,Jamil:2013nca,Herrera:2010vv,Rendall:2005if},
can produce a nearly scale invariant power spectrum, compatible with
the observational data in the context of $F(T)$ gravity. For a
relevant work on intermediate inflation in the context of $F(T)$
gravity, see \cite{Jamil:2013nca}. We shall investigate which $F(T)$
gravity can approximately realize the intermediate inflation
scenario, emphasizing at early cosmic times, and we shall calculate
the evolution of scalar perturbations and the corresponding spectral
index. As we shall demonstrate, the intermediate inflation scenario
in the context of $F(T)$ gravity, produces a nearly scale invariant
power spectrum, with a spectral index compatible with the latest
(2015) observational data, coming from the Planck collaboration
\cite{Ade:2015lrj}. In addition, we perform an analysis of the free
parameters space and we investigate for which values of the
parameters, compatibility with the Planck data can be achieved.

This paper is organized as follows: In section II we present in
brief some essential features of the $F(T)$ gravity and also the
necessary formalism for the sections to follow. In section III we
investigate which $F(T)$ gravity can approximately generate the
intermediate inflation scenario, emphasizing at early times, and we
calculate the power spectrum of primordial perturbations. In
addition we calculate the spectral index and we discuss its
compatibility with the observational data. In section IV we perform
an analysis of the free parameters space and we discuss how the
compatibility with the observational data can be achieved for
various values of the parameters. Finally, the conclusions follow at
the end of the paper.

\section{Essential Features of $F(T)$ Gravity}

For the $F(T)$ gravity formalism, the orthonormal tetrad components
$e_A (x^{\mu})$ are used, which are very common in teleparallelism
theories, where the index $A$ is $A=0,1,2,3$, on the tangent space
of each spacetime point $x^{\mu}$ of the spacetime manifold.
Effectively, the tetrads $e_A^\mu$ form the tangent vector of the
spacetime manifold. The tetrad components are related to the
spacetime metric $g^{\mu\nu}$  as follows $g_{\mu\nu}=\eta_{A B}
e^A_{\mu} e^B_{\nu}$. The torsion $T^\rho_{\verb| |\mu \nu}$ and the
contorsion $K^{\mu\nu}_{\verb||\rho}$ tensors, are defined as
follows, \begin{align} \label{eq:2.2}& T^\rho_{\ \mu\nu} \equiv
e^\rho_A \left(
\partial_\mu e^A_\nu -
\partial_\nu e^A_\mu \right)\,, \\ \notag & K^{\mu\nu}_{\ \ \rho} \equiv
-\frac{1}{2} \left(T^{\mu\nu}_{\ \ \rho} - T^{\nu \mu}_{\ \ \rho} -
T_\rho^{\ \ \mu\nu}\right)\,. \end{align} In addition, the torsion
scalar $T$ is defined as follows, \cite{Hayashi:1979qx} \be
\label{eq:2.4} T \equiv S_\rho^{\ \mu\nu} T^\rho_{\ \mu\nu}\,, \quad
S_\rho^{\ \mu\nu} \equiv \frac{1}{2} \left(K^{\mu\nu}_{\ \ \rho} +
\delta^\mu_\rho T^{\alpha \nu}_{\ \ \alpha}
 - \delta^\nu_\rho T^{\alpha \mu}_{\ \ \alpha}\right)\,.
\ee In effect, the  $F(T)$ modified teleparallel gravity action is
equal to, \be \label{eq:2.6} S = \int d^4x |e| \left[
\frac{F(T)}{2{\kappa}^2} +{\mathcal{L}}_{\mathrm{matter}} \right]\,
, \ee where $|e|= \det \left(e^A_\mu \right)=\sqrt{-g}$. By varying
the gravitational action of Eq.~(\ref{eq:2.6}), with respect to the
vierbein $e_A^\mu$, we get,
 \begin{align} \label{eq:2.7}
& \frac{1}{e}
\partial_\mu \left( eS_A^{\ \mu\nu} \right) F'
 - e_A^\lambda T^\rho_{\ \mu \lambda} S_\rho^{\ \nu\mu} F'
\\ \notag & +S_A^{\ \mu\nu} \partial_\mu T f'' +\frac{1}{4} e_A^\nu F =
\frac{{\kappa}^2}{2} e_A^\rho T_{(\mathrm{matter}\, \rho}^{\ \ \ \ \
\nu}\, . \end{align} By considering a flat FRW metric with line
element, \be \label{JGRG14} ds^2 = - dt^2 + a(t)^2 \sum_{i=1,2,3}
\left(dx^i\right)^2\, , \ee then we find that the tetrad components
are equal to, $e^A_\mu = \mathrm{diag} \left(1,a,a,a \right)$, which
implies that $g_{\mu \nu}= \mathrm{diag} \left(1, -a^2, -a^2, -a^2
\right)$. In effect, the torsion scalar $T$ is equal to, $T=-6H^2$,
and in addition, for the flat FRW metric, Eq.~(\ref{eq:2.7})
becomes,
 \be \label{FT} \frac{3}{\kappa^2} H^2 =
\rho_\mathrm{matter} + \rho_\mathrm{DE}\, , \quad \frac{1}{\kappa^2}
\left( H^2 + \dot H \right) = p_\mathrm{matter} + p_\mathrm{DE} \, ,
\ee where, \begin{align} \label{eq:4.3} & \rho_{\mathrm{DE}} =
\frac{1}{2{\kappa}^2} \left( -T -f +2T f' \right) \, , \\ \notag &
p_{\mathrm{DE}} = -\frac{1}{2{\kappa}^2} \left( \left(4 - 4 f' - 8 T
F''\right) \dot H
 -T -f +2T f' \right)\, ,
\end{align} and the prime denotes differentiation with respect to $T$. The
energy density $\rho_{\mathrm{DE}}$ and the pressure
$p_{\mathrm{DE}}$ appearing in Eq.~(\ref{eq:4.3}), satisfy the usual
continuity equation \be \label{eq:4.5} 0 =
\dot{\rho}_{\mathrm{DE}}+3H \left( \rho_{\mathrm{DE}} +
p_{\mathrm{DE}} \right) \, , \ee and the same applies for the energy
density $\rho_\mathrm{matter}$ and for the pressure
$p_\mathrm{matter} $, which correspond to the perfect matter fluids
which are present. In the purely vacuum case we have
$\rho_\mathrm{matter} = p_\mathrm{matter} = 0$, and the first
equation in Eq.~(\ref{FT}) becomes, \be \label{FT2} 0 = - F + 2 T F'
\, , \quad 0=F' + 2TF''\, , \ee where we used the fact that
$T=-6H^2$. In the following sections we shall use the above
formalism and we shall apply the above reconstruction technique in
order to find the $F(T)$ gravity which realizes the intermediate
inflation scenario.

\section{Intermediate Inflation and Evolution of Perturbations in $F(T)=T+f(T)$ Gravity}

As we already mentioned, in this paper we shall be interested in the
intermediate inflation scenario, which is quite popular in the
literature, see for example
\cite{Barrow:1990td,Barrow:1993zq,Rezazadeh:2014fwa,Barrow:2006dh,Barrow:2014fsa,Herrera:2014mca,Jamil:2013nca,Herrera:2010vv,Rendall:2005if}.
In the standard approach of intermediate inflation, it was worked
out in the context of scalar-tensor gravity
\cite{Barrow:1993zq,Rezazadeh:2014fwa,Barrow:2006dh,Barrow:2014fsa},
and its viability as a cosmological theory of the early Universe was
examined in Refs.
\cite{Barrow:2006dh,Barrow:2014fsa,Herrera:2014mca}. We shall
investigate whether this model of inflation is viable in the context
of $F(T)$ gravity. The intermediate inflation scale factor and the
Hubble rate are \cite{Barrow:1990td},
\begin{equation}\label{bambabounce}
a(t)=e^{A\,t^n},\,\,\, H(t)=A n t^{n-1}\, ,
\end{equation}
where $0<n<1$ and also $A>0$. The calculation of the evolution of
primordial perturbations in the context of $F(T)$ gravity, can be
found in various texts in the literature
\cite{Cai:2011tc,Chen:2010va,Izumi:2012qj,Nashed:2014lva,Hanafy:2014bsa,Hanafy:2014ica},
and we adopt the formalism and notation of Ref. \cite{Cai:2011tc}.
For a general approach applicable to various cosmological contexts,
see for example \cite{Brandenberger:2016vhg}. We shall be interested
in the longitudinal gauge, and therefore we consider only
scalar-type metric fluctuations. In effect, the perturbed metric has
the following form,
\begin{equation}\label{metricscalar}
\mathrm{d}s^2=(1+2\Phi)\mathrm{d}t^2-a(t)^2(1-2\Psi)\sum_i\mathrm{d}x^2_i\,
.
\end{equation}
It is conceivable that the scalar fluctuations of the metric are
quantified by the scalar functions $\Phi$ and $\Psi$. The leading
order perturbation of the torsion scalar can be expressed in terms
of the functions $\Phi$ and $\Psi$ as follows,
\begin{equation}\label{trosionpertubrscalar}
\delta T=12 H(\dot{\Phi}+H\Psi)\, ,
\end{equation}
where $H$ stands for the Hubble rate. In effect, by using the
formalism and the equations of motion of the previous section, we
find the following perturbation equations of the
$F(T)$\footnote{Note that we assumed $F(T)=T+f(T)$ for the $F(T)$
gravity entering the action.} gravity \cite{Cai:2011tc}, which
correspond to the perturbed metric (\ref{metricscalar}),
\begin{align}\label{ftgravieqnspertrb}
&
(1+f_{,T})\frac{\nabla^2}{a^2}\Psi-3(1+f_{,T})H\dot{\Psi}-3(1+f_{,T})H^2\Phi\\
\notag & +36f_{,TT}H^3(\dot{\Psi}+H\Phi)=4\pi G \delta \rho\, ,\\
\notag &
(1+f_{,T}-12H^2f_{,TT})(\dot{\Psi}+H\Phi)=4\pi G\delta q\, ,\\
\notag &(1+f_{,T})(\Psi-\Phi)=8\pi G\delta s\, , \\ \notag &
(1+f_{,T}-12H^2f_{,TT})\ddot{\Psi}+3H(1+f_{,T}
\\
\notag & -12H^2f_{,TT}-12\dot{H}f_{,TT}+48H^2\dot{H}f_{,TTT})\dot{\Psi}\\
\notag & +\Big{[}
3H^2(1+f_{,T}-12H^2f_{,TT})+2\dot{H}(1+f_{,T}-30H^2f_{,TT}\\
\notag &
+72H^4f_{,TTT})\Big{]}\Phi+\frac{1+f_{,T}}{2a^2}\nabla^2(\Psi-\Phi)=4\pi
G \delta p \, ,
\end{align}
with $f_{,T}$, standing for $\partial_T f(T)$, and the derivatives
$f_{,TT}$ and $f_{,TTT}$ can be found accordingly. Moreover, the
functions $\delta p$, $\delta \rho$, $\delta q$, $\delta s$,  denote
the fluctuations of the total pressure, of the total energy density
and of the fluid velocity and of the anisotropic stress
respectively. Assuming that a canonical scalar field represents the
matter fluid present, which has a potential $V(\phi)$, we get the
following equations,
\begin{align}\label{deltarhodrelations}
& \delta
\rho=\dot{\phi}(\delta\dot{\phi}-\dot{\phi}\Phi)+V_{,\phi}\delta
\phi\, ,\\ \notag & \delta q=\dot{\phi}\delta \phi\, ,\\ \notag &
\delta s=0\, ,\\ \notag & \delta p=\dot{\phi}(\delta
\dot{\phi}-\dot{\phi}\Phi)-V_{,\phi}\delta \phi\, .
\end{align}
In view of the above equations, it was shown in Ref.
\cite{Cai:2011tc}, that the relation $\Psi=\Phi$ holds true, and in
addition, the scalar fluctuation $\delta \phi$, uniquely determines
the gravitational potential $\Phi$. In effect, the $f(T)$ gravity
minimally coupled to a scalar field has one degree of freedom. With
regard to the evolution of the scalar perturbations, the following
equation determine how these evolve in time \cite{Cai:2011tc},
\begin{equation}\label{masterequation}
\ddot{\Phi}_k+\alpha
\dot{\Phi}_k+\mu^2\Phi_k+c_s^2\frac{k^2}{a^2}\Phi_k=0\, ,
\end{equation}
with $\Phi_k$ being the scalar Fourier mode of the potential $\Phi$,
and in addition, the functions $c_s^2$, $\alpha$ and $\mu^2$ are the
the speed of sound parameter, the frictional term and the effective
mass respectively, corresponding to the scalar potential $\Phi$. In
detail, the latter functions are equal to,
\begin{align}\label{functionsanalytical}
&
\alpha=7H+\frac{2V_{,\phi}}{\dot{\phi}}-\frac{36H\dot{H}(f_{,TT}-4H^2f_{,TTT})}{1+f_{,T}-12H^2f_{,TT}}\,
,\\ \notag &
\mu^2=6H^2+2\dot{H}+\frac{2HV_{,\phi}}{\dot{\phi}}-\frac{36H^2\dot{H}(f_{,TT}-4H^2f_{,TTT})}{1+f_{,T}-12H^2f_{,TT}}\,
, \\ \notag & c_s^2=\frac{1+f_{,T}}{1+f_{,T}-12H^2f_{,TT}}\, .
\end{align}
The equation of motion for the canonical scalar field is,
\begin{equation}\label{eqnforscaux}
\ddot{\phi}+3H\dot{\phi}+V_{,\phi}=0\, ,
\end{equation}
and thus by rewriting the $f(T)$ gravity equation of motion as
follows,
\begin{equation}\label{secondfriedmann}
(a+f_{,T}-12H^2f_{,TT})\dot{H}=-4\pi G \dot{\phi}^2\, ,
\end{equation}
the master equation that determines the evolution of scalar
perturbations is,
\begin{equation}\label{evolutionequationfinal}
\ddot{\Phi}_k+\left(H-\frac{\ddot{H}}{\dot{H}}\right)\dot{\Phi}_k+\left(2\dot{H}-\frac{H\ddot{H}}{\dot{H}}
\right)\Phi_k+\frac{c_s^2k^2}{a^2}\Phi_k=0\, .
\end{equation}
As it can be seen from the structure of Eq.
(\ref{evolutionequationfinal}), it is identical to the
Einstein-Hilbert master equation, apart from the appearance of the
speed of sound parameter.

A physical quantity that can adequately quantify any cosmological
inhomogeneities, is the comoving curvature fluctuation, which we
denote as $\zeta$, which in the case at hand is equal to,
\begin{equation}\label{comovcurv}
\zeta =\Phi-\frac{H}{\dot{H}}\left (\dot{\Phi}+H\Phi \right)\, .
\end{equation}
The comoving curvature fluctuation is gauge invariant and it
simplifies significantly the calculation of the spectral index. We
also introduce the quantity $v$ which is defined as follows,
\begin{equation}\label{varintr1}
v=z\zeta\, ,
\end{equation}
where $z$ stands for,
\begin{equation}\label{fgr}
z=a\sqrt{2\epsilon}\, ,
\end{equation}
and also $\epsilon$ is the first slow-roll index
$\epsilon=-\frac{\dot{H}}{H^2}$. In terms of the new variables $v$
and $z$, the master equation that determines the evolution of
primordial curvature perturbations becomes \cite{Cai:2011tc},
\begin{equation}\label{mastereqn2}
v_k''+\left( c_s^2k^2-\frac{z''}{z}\right) v_k=0\, ,
\end{equation}
where the sound speed parameter appears in Eq.
(\ref{functionsanalytical}). Note here that the ``prime'' in Eq.
(\ref{mastereqn2}) denotes differentiation with respect to the
conformal time $\tau$, which in terms of the cosmic time $t$ is
defined as follows,
\begin{equation}\label{conformtime}
\tau=\int \mathrm{d}t\frac{1}{a}\, .
\end{equation}
We need to note that the $f(T)$ gravity affects the evolution of the
primordial curvature perturbations (\ref{mastereqn2}) via the speed
of sound parameter $c_s$, and for the flat FRW background of Eq.
(\ref{JGRG14}), the $f(T)$ gravity first Friedmann equation becomes,
\begin{equation}\label{fteqn1}
H^2=-\frac{f(T)}{6}-2f_{,T}H^2\, ,
\end{equation}
and since $T=-6H^2$, for the intermediate inflation case
(\ref{bambabounce}) we obtain,
\begin{equation}\label{explicitttrelation}
T=-6 A^2 n^2 t^{2 n-2}\, .
\end{equation}
By solving the above equation with respect to the cosmic time $t$,
we obtain the function $t(T)$, which is,
\begin{equation}\label{ttexplic}
t(T)=6^{-\frac{1}{2 n-2}} \left(-\frac{T}{A^2
n^2}\right)^{\frac{1}{2 n-2}}\, .
\end{equation}
By combining Eqs. (\ref{fteqn1}) and (\ref{ttexplic}), we obtain the
approximate form of the $f(T)$ gravity which realizes the
intermediate inflation scenario, which is,
\begin{equation}\label{ftgravitytyfn1}
f(T)=c_1 T^{\frac{A n}{2}}-\frac{T}{2 \left(1-\frac{A
n}{2}\right)}\, ,
\end{equation}
with $c_1$ being an arbitrary integration constant, and hence the
total $F(T)$ gravity is $F(T)=T+f(T)$. At this point we shall
express all the quantities as function of the conformal time $\tau$,
however in the case at hand, certain simplifications can be made.
Particularly, since we are interested for the inflationary era, this
means that we are interested in the early-time era, so the cosmic
time variable takes small values. In effect, the exponential
$e^{At^n}$ for small values of $t$ can be approximated as
$e^{At^n}\sim 1$, and therefore we may identify the cosmic time $t$
with the conformal time $\tau$ (see relation (\ref{conformtime})),
that is $\tau \sim t$. In effect, the function $z(t)$ for the
intermediate $f(T)$ inflation reads,
\begin{equation}\label{zrt}
z(t)=\sqrt{\frac{2 (1-n) t^{-n}}{A n}}\, ,
\end{equation}
and moreover, by using Eq. (\ref{ftgravitytyfn1}) the sound speed
parameter is equal to,
\begin{align}\label{cscalculation}
& c_s^2(t)=\frac{A \,c_1\, n 6^{\frac{A n}{2}} (A n-2) \left(-A^2
n^2 t^{2 n-2}\right)^{\frac{A n}{2}}}{2 (A n-1) S(t) }\\ \notag &
-\frac{12 A^2 n^2 (A n-1) t^{2 n-2}}{2 (A n-1) S(t)} \, ,
\end{align}
where $S(t)$ is,
\begin{align}\label{akyroeqns}
& S(t)=\Big{(}c_1 6^{\frac{A n}{2}} (A n-2) \left(-A^2 n^2 t^{2
n-2}\right)^{\frac{A n}{2}}\\
\notag & -6 A^2 n^2 t^{2 n-2}\Big{)}\, .
\end{align}
Due to the fact that the dominant term in the expression of Eq.
(\ref{cscalculation}) is $\sim t^{2 n-2}$, we may approximate the
sound speed parameter as $c_s^2\simeq 1$. In effect, the master
equation describing the evolution of the perturbations can be
written as follows,
\begin{equation}\label{mastereqn23}
v''_k(t)+\left( k^2-\frac{\left(\frac{n}{2}+1\right) n}{2
t^2}\right) v_k(t)=0\, ,
\end{equation}
which has the following solution,
\begin{equation}\label{evoltpertbsol}
v_k(t)=C_1\sqrt{t}J_{\frac{n+1}{2}}(k t)+C_2\sqrt{t}
Y_{\frac{n+1}{2}}(kt)\, ,
\end{equation}
with $J_n(z)$ and $Y_n(z)$ being the Bessel functions of first and
second kind respectively, and in addition $C_1$ and $C_2$ are
arbitrary integration constants. The expression of Eq.
(\ref{evoltpertbsol}) can be simplified for small values of the
Bessel functions arguments, so we get the simplified result for
$v_k(t)$, which is,
\begin{align}\label{evoltpertbsol12}
& v_k(t)= \frac{\text{C1} \sqrt{t} (k t)^{n/2} \left(2^{\frac{1}{2}
(-n-1)} \sqrt{k t}\right)}{\Gamma \left(\frac{n+1}{2}+1\right)} \\
\notag &  +\text{C2}\sqrt{t}
\Big{(}-\frac{2^{\frac{n}{2}+\frac{1}{2}} \Gamma
\left(\frac{n+1}{2}\right) (k t)^{-\frac{n}{2}-\frac{1}{2}}}{\pi
}\\
\notag &-\frac{2^{-\frac{n}{2} -\frac{1}{2}} \cos \left(\frac{1}{2}
\pi (n+1)\right) \Gamma \left(\frac{1}{2} (-n-1)\right) (k
t)^{\frac{n}{2}+\frac{1}{2}}}{\pi }\Big{)}\, ,
\end{align}
and by keeping only leading order terms we acquire,
\begin{equation}\label{dominanthubbleev}
v_k(t)=\frac{C_2 \sqrt{t} \left(2^{\frac{n}{2}+\frac{1}{2}} \Gamma
\left(\frac{n+1}{2}\right) (k
t)^{-\frac{n}{2}-\frac{1}{2}}\right)}{\pi }\, .
\end{equation}
At this point we proceed in calculating the power spectrum of the
primordial curvature perturbations, which in terms of the functions
$v_k$ and $z$ is defined as follows,
\begin{equation}\label{powerspectrmrelation}
\mathcal{P}_{\zeta}=\frac{k^3}{2\pi^2}|\frac{v_k}{z}|_{k=a H}^2\, ,
\end{equation}
and note that it must be evaluated at the horizon crossing, which
occurs when $k=aH$, where $k$ is the wavenumber of each primordial
mode. In order to find an analytical expression for the power
spectrum, we need to express the quantity $|\frac{v_k}{z}|^2$ as a
function of the wavenumber $k$. Note that, firstly, we already have
the function $v_k(t)$, which we calculated the resulting expression
in Eq. (\ref{dominanthubbleev}), however, the parameter $C_2$ is
also $k$-dependent. This $k$-dependence of $C_2$ can be determined
by assuming some initial conditions for the function $v_k(t_0)$.
Particularly we assume that it originates from a Bunch-Davies
vacuum, and hence we have $v_k\simeq \frac{e^{-ikt}}{\sqrt{2k}}$. In
effect the constant $C_2$ as a function of the wavenumber is,
\begin{equation}\label{constantc4}
C_4\simeq \frac{\pi  2^{-\frac{n}{2}-\frac{1}{2}} k^{n/2}
t^{n/2}}{\Gamma \left(\frac{n+1}{2}\right)}\, .
\end{equation}
We need to note that the assumption of the Bunch-Davies vacuum is an
assumption based on the fact that the intermediate inflation
scenario is an inflationary scenario. The Bunch-Davies vacuum
assumption is based on the fact that the initial state corresponds
to an era that the curvature does not affect the fluctuations. It is
possible though that before the inflationary era, another scenario
might occur, like for example a bouncing phase \cite{ref1}, so in
effect the Bunch-Davies assumption would be inapplicable. Therefore,
one should adopt the approach of Refs. \cite{ref12,ref2} in order to
find the correct approximation for the initial state. In fact, as it
shown in \cite{ref3}, in some cases the intermediate inflation
scenario is identical with the expanding phase of a singular bounce.
Then, the considerations of Refs. \cite{ref12,ref2} might be
compelling. However, for the purposes of this paper we assume that
the initial state remains a Bunch-Davies vacuum state and we defer
this non-trivial task to a future work.

Another quantity that has an implicit $k$-dependence is the cosmic
time in both Eqs. (\ref{dominanthubbleev}) and (\ref{constantc4}).
In order to find the $k$-dependence, recall that the power spectrum
is evaluated at the horizon crossing, where the equation $k=aH$, and
since $a\sim 1$ at early times, we have,
\begin{equation}\label{inthorcross}
t\simeq \frac{k^{\frac{1}{n-1}}}{(A n)^{\frac{1}{n-1}}}\, .
\end{equation}
Hence, by combining Eqs. (\ref{zrt}), (\ref{dominanthubbleev}),
(\ref{constantc4}) and (\ref{inthorcross}), the resulting power
spectrum of Eq. (\ref{powerspectrmrelation}) is equal to,
\begin{equation}\label{powerspectrumfinalrelationforft}
\mathcal{P}_{\zeta}\simeq \frac{\left(n^{1-\frac{n}{n-1}}
A^{1-\frac{n}{n-1}}\right) k^{\frac{1}{n-1}+3}}{4 \pi ^2 (1-n)}\, .
\end{equation}
Obviously, the power spectrum is not scale invariant, however, as we
show in the next section, it can be compatible with the Planck data,
since a nearly scale invariant spectrum is produced for specifically
chosen values of the parameter $n$.

\subsection{Comparison with Planck 2015 Data and Analysis of the Parameter Space}

Having the expression for the power spectrum of primordial curvature
perturbations at hand, namely Eq.
(\ref{powerspectrumfinalrelationforft}), we can calculate the
spectral index straightforwardly, by using the following relation,
\begin{equation}\label{spcrelation}
n_s-1=\frac{\mathrm{d}\ln \mathcal{P}_{\zeta}}{\mathrm{d}\ln k}\, .
\end{equation}
Therefore, for the power spectrum of Eq.
(\ref{powerspectrumfinalrelationforft}), the spectral index reads,
\begin{equation}\label{nsfinalforft}
n_s=\frac{1}{n-1}+4\, ,
\end{equation}
and now we investigate whether this spectral index can be compatible
with the 2015 observational data of the Planck collaboration
\cite{Ade:2015lrj}, in which case the spectral index is constrained
as follows,
\begin{equation}
\label{planckdata} n_s=0.9644\pm 0.0049\, .
\end{equation}
In effect, a spectral index $n_s$ that takes values in the interval
$n_s=[0.9595,0.9693]$ can be considered compatible with the Planck
constraints (\ref{planckdata}).
\begin{figure}[h] \centering
\includegraphics[width=18pc]{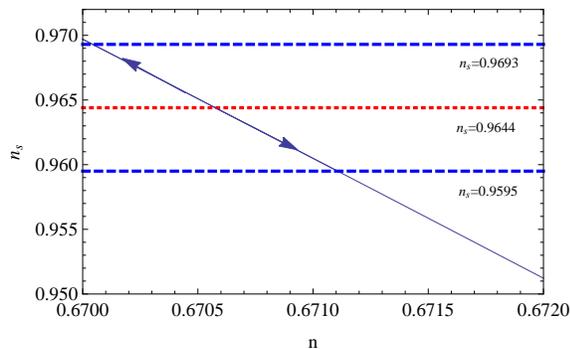}
\caption{The spectral index $n_s=\frac{1}{n-1}+4$ as a function of
the parameter $n$. The curve with arrows indicates the values of $n$
for which the spectral index is compatible with the Planck 2015
constraints on the spectral index. }\label{plota}
\end{figure}
In Fig. \ref{plota}, we have plotted the spectral index
$n_s=\frac{1}{n-1}+4$, as a function of $n$, and as it can be seen,
the allowed range of $n$ is indicated by the arrows. The upper
dashed line (red in online version) indicates the highest value of
the spectral index that it is allowed, namely $n_s=0.9693$, while
the lower dashed line (red in online version) indicates the lowest
allowed value of the spectral index, namely $n_s=0.9595$. The middle
dashed (red in online version) line corresponds to $n_s=0.9644$.
Hence if $n$ is chosen to take values in the range
$n=[0.67,0.6711]$, then the spectral index $n_s$ takes values in the
interval $n_s=[0.9595,0.9693]$, and hence it is compatible with the
Planck data. In conclusion, for the $F(T)$ gravity description of
the intermediate inflation scenario, the resulting spectral index
depends only on one parameter $n$, that appears in the scale factor
of the intermediate inflation scenario, and it can be compatible
with the Planck 2015 observational data, if $n$ takes values in the
interval $n=[0.67,0.6711]$.

\subsection{Discussion on the Scalar-to-tensor Ratio}

We now discuss the calculation of the scalar-to-tensor ratio, which
in the case at hand it will prove a non-trivial task. Consider the
perturbation of the flat FRW metric,
\begin{equation}\label{flatperturbation}
\mathrm{d}s^2=a(\tau)^2\left(
-\mathrm{d}\tau^2+(\delta_{ij}+h_{ij}\mathrm{d}x^i\mathrm{d}x^j)\right)\,
,
\end{equation}
where the tensor perturbation $h_{ij}$ is real, transverse and
traceless, that is, $h_{ij}=h_{ji}$, $h_{ii}=0$ and $h_{ij,j}=0$.
The calculation of the amplitude of the tensor modes can be done by
using standard approaches, see for example \cite{Chen:2010va,ref4},
so the resulting master differential equation that governs the
evolution of the tensor perturbations is,
\begin{equation}\label{masterdiffeqntensormodes}
\left(\ddot{h}_{ij}+3H\dot{h}_{ij}-\frac{\nabla^2}{a^2}h_{ij}
\right)-\frac{12H\dot{H}f_{,TT}}{1+f_{,T}}\dot{h}_{ij}=0\, .
\end{equation}
By expanding the tensor perturbation $h_{ij}$ in Fourier modes, one
can calculate the evolution of each tensor Fourier mode. However,
for the case at hand this is a highly non-trivial task, and it is
very difficult to obtain analytic results even by taking the small
$t$ limit. This is due to the presence of the last term in Eq.
(\ref{masterdiffeqntensormodes}), which contains the derivatives of
the function $f(T)$, which in our case, by using Eq.
(\ref{ftgravitytyfn1}), the last term reads,
\begin{align}\label{ftcaseredas}
&-\frac{12H\dot{H}f_{,TT}}{1+f_{,T}}\dot{h}_{ij}= \\ \notag & \left(
\frac{6 A^3 C_1 (n-1) n^3 (A n-2)^2 t^{2 n-3} T^{\frac{A n}{2}-1}}{A
C_1 n (A n-2) T^{\frac{A n}{2}}+2 T (A n-1)}\right)\dot{h}_{ij}\, ,
\end{align}
and by using Eq. (\ref{explicitttrelation}), we can express it in
terms of the cosmic time (recall that the cosmic and the conformal
time are equivalent in the small $t$ limit in our case, see Eq.
(\ref{conformtime}), with $a\sim 1$),
\begin{align}\label{ftcaseredas1}
& -\frac{12H\dot{H}f_{,TT}}{1+f_{,T}}\dot{h}_{ij}=\\ \notag & \left(
\frac{(n-1) t (A n-2)^2}{\frac{A n 2^{2-\frac{A n}{2}} 3^{1-\frac{A
n}{2}} (A n-1) t^{2 n} \left(A^2 n^2 t^{2 n-2}\right)^{-\frac{1}{2}
(A n)}}{C_1}+t^2 (2-A n)}\right)\dot{h}_{ij}\, .
\end{align}
Hence the study of the tensor modes is highly non-trivial and we
hope that we address this is issue in a future work focused solely
on the calculation of the evolution of the tensor modes.

\section{Conclusions}

In this paper we investigated how the intermediate inflation
scenario can be realized by an $F(T)$ gravity, and we calculated the
power spectrum of the primordial curvature perturbations. The
original question was whether the spectral index of the intermediate
inflation scenario realized by $F(T)$ gravity, can be compatible
with the 2015 Planck constraints, and as we showed, it is possible
and for a quite large range of the values of the free parameters.
Actually, as we showed, the power spectrum depends on the parameter
$n$, which appears in the scale factor $a\sim e^{A\,t^n}$ of the
intermediate inflation scenario, and when $n$ takes its values in
the interval $n=[0.67,0.6711]$, the resulting spectral index is
compatible with the Planck data.

Before we close, we need to highlight an observation with regard to
the intermediate inflation scenario. By looking at functional form
of the scale factor and of the Hubble rate of the intermediate
inflation scenario, it is easy to realize that the intermediate
inflation scenario has a finite time singularity at $t=0$, and since
$0<n<1$, by following the classification of Refs.
\cite{Nojiri:2005sx,Oikonomou:2015qfh,Kleidis:2016vmd}, it is easy
to see that the singularity is a Type II singularity. In effect, it
is softer than the Big Bang singularity, since it is simply a
pressure singularity. Also we need to stress that the intermediate
inflation scenario could be identified with the expanding phase of
the so-called singular bounce
\cite{Odintsov:2015ynk,Oikonomou:2015qha}, but with the singularity
occurring at the origin being a Type II, instead of a Type IV which
was the case in Refs. \cite{Odintsov:2015ynk,Oikonomou:2015qha}. An
interesting scenario could be that before the intermediate inflation
scenario occurs, a contracting bouncing phase occurs, like in bounce
inflation scenarios \cite{Piao:2003zm,Piao:2005ag,Saidov:2010wx}.
This scenario could be realized in the context of modified gravity,
but we defer this project to a future work.

\section*{Acknowledgments}

This work is supported by Min. of Education and Science of Russia
(V.K.O).

\end{document}